\documentclass[twocolumn,10pt]{revtex4-1}
\def \and{\& }

\def\by#1{#1,}
\def\and{and }
\def\yr#1{{(#1)}}
\def\paper#1{#1}
\def\jour#1{{\it #1}}

\def\vol#1{{\bf #1},}
\def\issue#1{}
\def\pages#1{\hbox{#1},}

\usepackage{amsmath}
\usepackage{amsfonts}
\usepackage{amssymb}
\usepackage{graphicx}
\usepackage{appendix}
\usepackage{placeins}

\begin{document}

\title{Surface permeability, capillary transport and the Laplace-Beltrami problem.}
\author{Penpark Sirimark, Alex V. Lukyanov \& Tristan Pryer }
\affiliation{School of Mathematical and Physical Sciences, University of Reading, Reading RG6 6AX, UK}

\begin{abstract}
We have established previously, in a lead-in study, that the spreading of liquids in particulate porous media at low saturation levels, characteristically less than 10\% of the void space, has very distinctive features in comparison to that at higher saturation levels. In particular, we have found that the dispersion process can be accurately described by a special class of partial differential equations, the super-fast non-linear diffusion equation. The results of mathematical modelling have demonstrated very good agreement with experimental observations. However, any enhancement of the accuracy and predictive power of the model, keeping in mind practical applications, requires the knowledge of the effective surface permeability of the constituent particles, which defines the global, macroscopic permeability of the particulate media.  In the paper, we demonstrate how this quantity can be determined through the solution of the Laplace-Beltrami Dirichlet problem, we study this using the well-developed surface finite element method.           
\end{abstract}

\maketitle

\section{Introduction}

Liquid distributions and transport in particulate porous media, such as sand, at low saturation levels $s$, defined in our study as the ratio of the liquid volume $V_L$ to the volume of available voids $V_E$ in a sample volume element $V$, $s=\frac{V_L}{V_E}$, have many distinctive features. Theoretically, as we have shown previously, the liquid dispersion can be described by a special class of mathematical models, the superfast non-linear diffusion equation~\cite{Lukyanov2012}.  Unlike in the standard porous medium equation, which is a paradigm of research in porous media~\cite{Vazquez-Book}, in this special case, the non-linear coefficient of diffusion $D(s)$ demonstrates divergent behaviour as a function of saturation $s$, $D(s)\propto (s-s_0)^{-3/2}$, where $s_0$ is some minimal saturation level~\cite{Lukyanov2012}.

In practical applications, the analysis of this regime of wetting is crucial for studies of biological processes, such as microbial activity, and spreading of persistent (non-volatile) liquids in soil compositions and dry porous media commonly found in arid natural environments and industrial installations~\cite{Lukyanov2012, Tuller-2005}. 

If we consider liquid distributions on the grain size length scale, one would observe that when the saturation level $s$ is reduced to (or below) the critical level $s_c\approx 10\%$, the liquid domain predominantly consists of isolated liquid bridges formed at the point of particle contacts~\cite{Lukyanov2012, Orr-Scriven-1975, Willett-2000, Herminghaus-2005, Herminghaus-2008, Herminghaus-2008-2}, see Fig. \ref{Fig1} for illustration. The formation of liquid bridges is characteristic for the so-called pendular regime of wetting~\cite{Orr-Scriven-1975, Willett-2000, Herminghaus-2005, Herminghaus-2008, Herminghaus-2008-2, Bear-Book}. In this regime, the liquid bridges are only connected via thin films formed on the rough particle surfaces and they serve as variable volume reservoirs, where the capillary pressure $p$ depends directly on the amount of the liquid in the bridge $V_b$  
\begin{equation}
\label{Pressure}
p\approx -p_0\left(\frac{R^3}{V_b}\right)^{1/2}.
\end{equation}
Here, $p_0=\frac{2\gamma}{R}\cos\phi_c$, $\gamma$ is the coefficient of the surface tension of the liquid, $\phi_c$ is the contact angle made by the free surface of the liquid bridge with the rough solid surface of the constituent particles and $R$ is an average radius of the porous medium particles~\cite{Orr-Scriven-1975, Herminghaus-2005, Lukyanov2012}. The spreading process in such conditions only occurs over the rough surface of the elements of the particulate porous media connecting the liquid bridges, Fig. \ref{Fig1}.

\section{Macroscopic formulation of the super-fast diffusion problem}
Microscopically, the liquid creeping flow through the surface roughness of each particle can be described by a local Darcy-like relationship~\cite{Yost-1998} between the surface flux density ${\bf q}$ and averaged (over some area containing many surface irregularities) pressure in the grooves $\psi$
\begin{equation}
\label{Macroscopic-Darcy-2}
-\frac{\kappa_m}{\mu}  \nabla \psi ={\bf q}. 
\end{equation} 
Here, $\mu$ is liquid viscosity and $k_m$ is the local coefficient of permeability of the rough surface, which proportional to the average amplitude of the
surface roughness $\delta_R$, that is the width of the surface layer conducting the liquid flux, $k_m\propto \delta_R^2$~\cite{Yost-1998}. We note that, if the rough surface layer is not fully saturated with the liquid, parameter $\delta_R$ should be interpreted as the characteristic width of the liquid layer within the rough surface layer. It is always assumed that $\delta_R\ll R$, that is the amplitude of the surface roughness (or the width of the liquid layer) is always much smaller than the particle size.

\begin{figure}[ht!]
\begin{center}
\includegraphics[trim=-1.5cm 3.cm 1cm -0.5cm,width=\columnwidth]{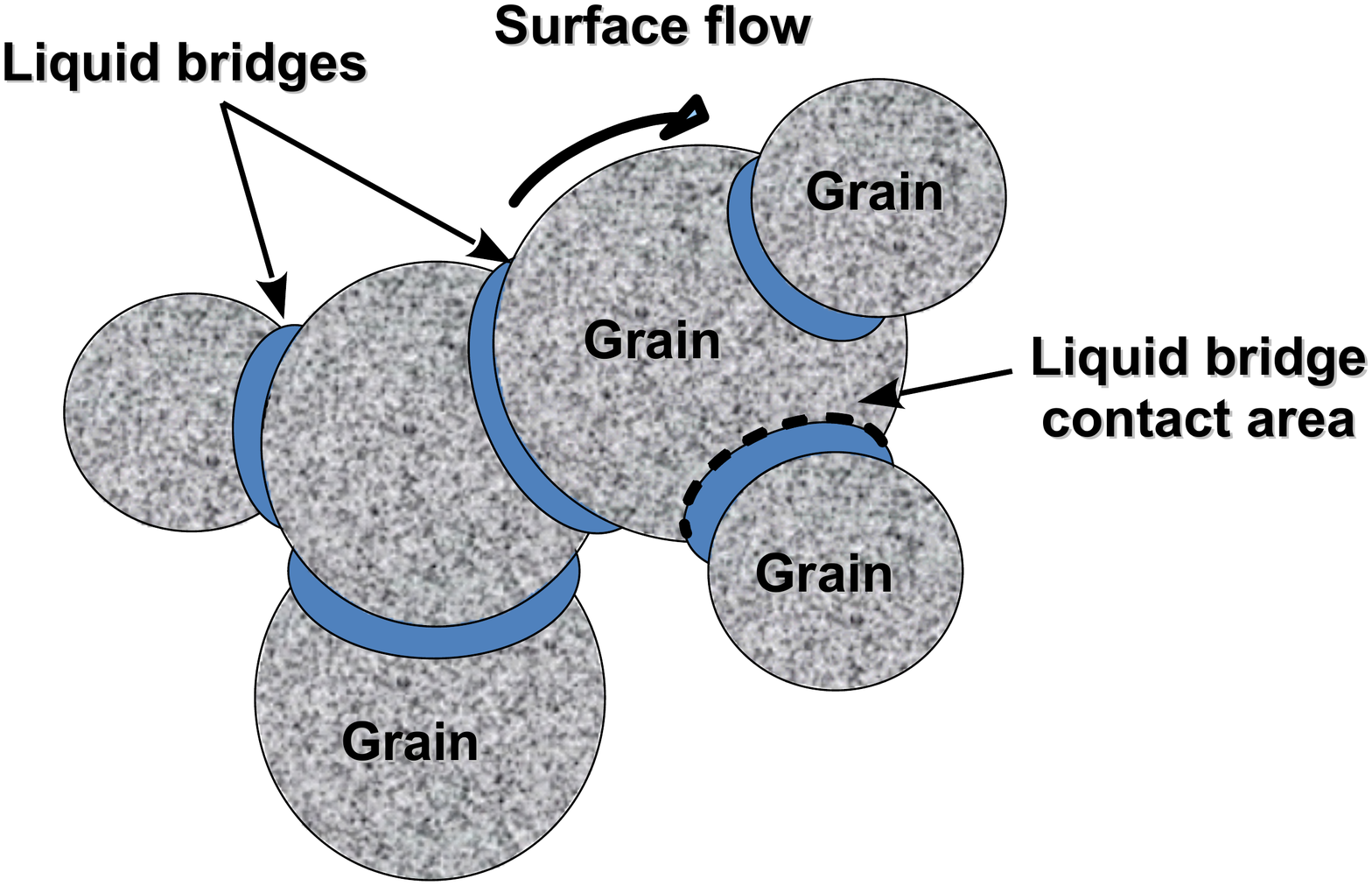}
\end{center}
\caption{Illustration of the liquid distribution in particulate porous media (grey) with pendular rings (blue) at low saturation levels.} 
\label{Fig1}
\end{figure}

Macroscopically, that is after averaging over some volume element containing many particles of the porous medium, the diffusion process in the slow creeping flow conditions can be described by a non-linear super-fast diffusion equation
\begin{equation}
\label{Superfast-1}
\frac{\partial s}{\partial t}= \nabla\cdot \left\{ D(s) \nabla s \right\},  \quad t>0,
\end{equation}
$$
D(s)= \frac{D_0}{(s-s_0)^{3/2}}, \quad s>s_0, 
$$
which directly follows from the conservation of mass principle
\begin{equation}
\label{mass-conservation}
\frac{\partial (\phi s)}{\partial t} + \nabla\cdot {\bf Q}=0.
\end{equation}
Here $D_0$ is the effective, macroscopic coefficient of non-linear diffusion, $s_0$ is the minimal level of saturation, which can be only achieved when the liquid bridges cease to exist ($s_0\approx 0.5\%$, see details in~\cite{Lukyanov2012, Herminghaus-2008, Herminghaus-2008-2}), $\phi$ is porosity defined as $\displaystyle \phi=\frac{V_E}{V}$, which is further assumed to be constant, and $\bf Q$ is the macroscopic flux density. The macroscopic flux density $\bf Q$ is defined in such a way that the total flux through the surface of a macroscopic sample volume element is given by the surface integral $\int {\bf Q\cdot n}\, dS$, where $\bf n$ is the normal vector to the surface of the element.

Equation (\ref{Superfast-1}) can be obtained from (\ref{mass-conservation}) using (\ref{Pressure})-(\ref{Macroscopic-Darcy-2}) and the spatial averaging theorem formulated in~\cite{Whitaker-1969} assuming that \cite{Lukyanov2012}:
\begin{itemize}
\item the rough surface area of the porous media particles is fully saturated with the liquid;
\item the liquid is incompressible;
\item the local Darcy's law (\ref{Macroscopic-Darcy-2}) is observed on the rough particle surface elements.
\end{itemize} 
All three criteria are usually very well satisfied in practical applications, and we will further assume that this is the case. The approximation of the fully saturated rough surface layer is well fulfilled, if the characteristic pressure amplitude $|\psi|$ is less than the capillary pressure amplitude defined on the length scale of the surface roughness $\delta_R$, which is of the order of $\delta_R\sim 1\,\mu\mbox{m}$ in typical sands \cite{Alshibli2004}, as is demonstrated in~\cite{Yost-1998}. That is, $|\psi| < \frac{\gamma}{\delta_R}$, and, for example for water ($\gamma=72\, \mbox{mN}/\mbox{m}$) at $\delta_R =1\,\mu\mbox{m}$, this results in $|\psi| < 7.2\times 10^4\, \mbox{Pa}$. Otherwise, at larger absolute values of the (negative) capillary pressure, the liquid volume within the surface roughness layer would start to vary leading to variations of the effective liquid surface layer thickness $\delta_R$, though, it is not difficult to introduce a correction~\cite{Ransohoff1988, Tokunaga-1997, Tuller-2000, Tuller-2005}. Note, in the formulation (\ref{Superfast-1}), the effects of gravity were neglected assuming that the capillary length $l_c=\sqrt{\gamma/\rho g_0}$ is much larger than the length scale associated with the gradient of the capillary pressure, that is $l_c\gg \sqrt{\delta_R L_0}$, where $L_0$ is the characteristic length scale of the wetting area. Here $g_0$ is the Earth gravity constant and $\rho$ is the liquid density, so that for most liquids $l_c\sim 1\,\mbox{mm}$. At the same time, taking $\delta_R\approx 1\,\mu\mbox{m}$ and $L_0\approx 10\,\mbox{mm}$, as it was in the experiments reported in~\cite{Lukyanov2012}, one gets $\sqrt{\delta_R L_0}\approx 0.1\,\mbox{mm}\ll l_c$.  

The effective coefficient of diffusion $D_0=f_{\phi}\frac{K}{\mu}$ comprises of the global permeability of the surface elements $K=k_m\frac{S_e}{S}$~\cite{Lukyanov2012}. Here, parameter $f_{\phi}=\frac{p_0}{2\phi}\sqrt{\frac{3 N_c}{4\pi} \frac{1-\phi}{\phi}}$, $N_c$ is a coordination number of the particles, that is the average number of contacts per a particle (in sands, typically, $N_c\approx 7$) and $S_e/S$ is the ratio of the effective area of entrances and exits of the liquid flow in a sample volume element with surface area $S$, see details in~\cite{Lukyanov2012}. Note, that the ratio $S_e/S$ is defined in such a way, that the microscopic flux density $\bf q$ averaged over the liquid volume $V_l$ within a macroscopic sample volume element $V$, $\langle{\bf q}\rangle^l=\int_{V_l}\, {\bf q}\, dV$, if multiplied by the ratio $\langle{\bf q}\rangle^l\frac{S_e}{S}=\bf{Q}$, would result in the macroscopic average flux density $\bf Q$. 

The global surface permeability of the particles $K$ is one of the main elements of the model that enables an accurate representation of the liquid dispersion at low saturation levels.  On the other hand, this quantity is difficult to accurately estimate a priori. It is fully defined by the particle shape and the dimension of the liquid bridge contact area, Fig. \ref{Fig1}. In this paper, we determine this important parameter on the basis of a solution to the Laplace-Beltrami problem in a representative case of a spherical (or nearly spherical) particle, which provides, as we will show, a reasonable approximation for the constituent elements of particulate porous media, such as sands.

\section{Microscopic model of the surface permeability of the elements.}

\begin{figure}[ht!]
\begin{center}
\includegraphics[trim=-1.5cm 3.cm 1cm -0.5cm,width=\columnwidth]{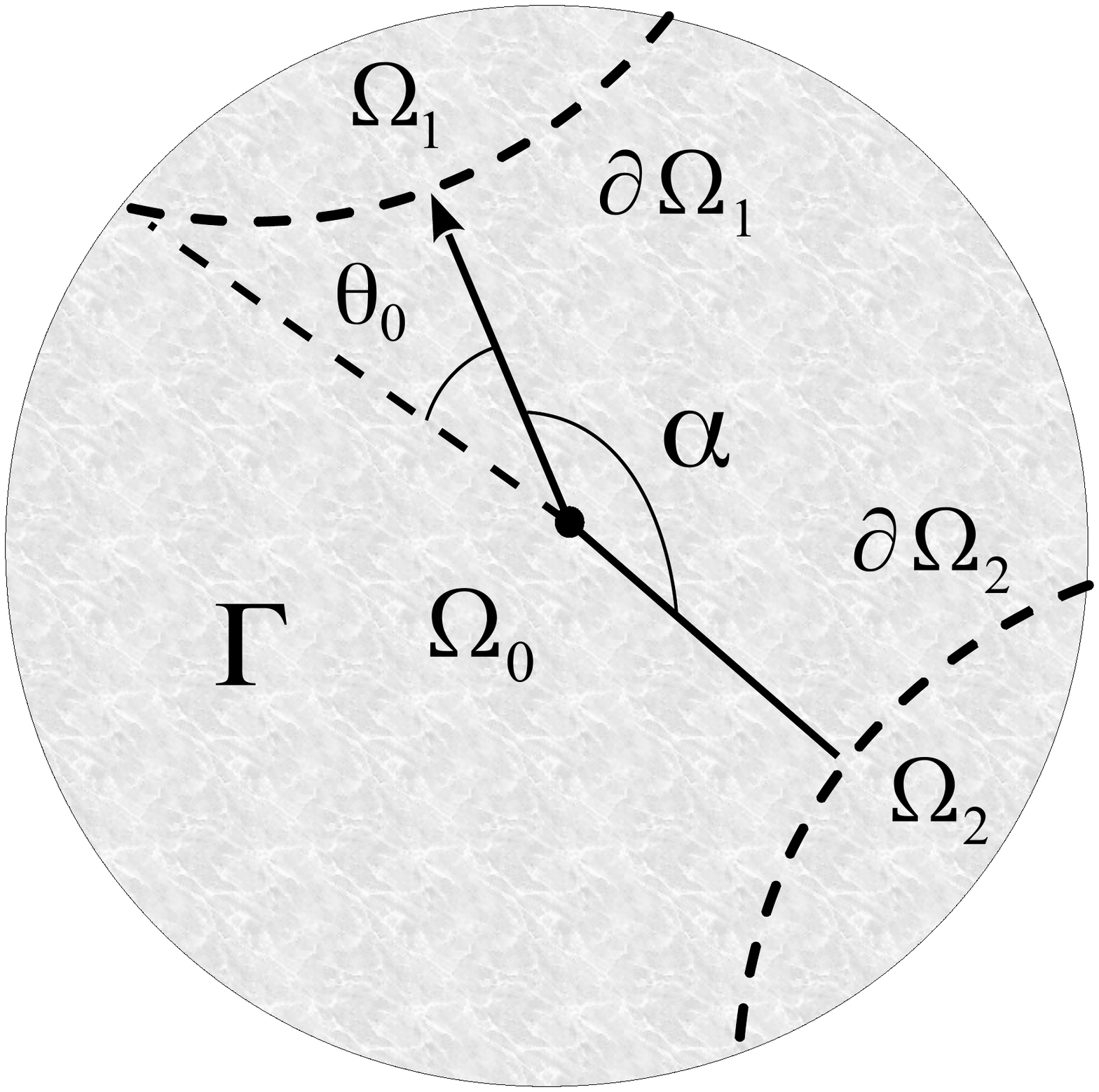}
\end{center}
\caption{Illustration of the solution domains on a spherical particle.} 
\label{Fig2}
\end{figure}

Consider, as the simplest example, a spherical particle of radius $R$ with a closed surface $\Gamma$, which is split into three sub-domains $\Omega_0$, $\Omega_1$ and $\Omega_2$ with the surface boundaries between them $\partial \Omega_1$ and $\partial \Omega_2$, as is shown in Fig. \ref{Fig2}. The location of the sub-domains $\Omega_1$ and $\Omega_2$ to each other on the surface is fixed by the tilt angle $\alpha$. The sub-domains $\Omega_1$ and $\Omega_2$ correspond to the contact area covered by the liquid in the bridges, while the surface flow, described by (\ref{Macroscopic-Darcy-2}), takes place in $\Omega_0$.

Since the rough surface area of the particles is assumed to be fully saturated in creeping flow conditions~\cite{Yost-1998}, liquid pressure $\psi$, due to incompressibility of the liquid, should satisfy the Laplace-Beltrami equation defined on the surface of the sub-domain $\Omega_0$
\begin{equation}
\label{Laplace-Beltrami}
\Delta_{\Omega_0} \psi=0,
\end{equation}   
as it follows from (\ref{Macroscopic-Darcy-2}). Here, $\Delta_{\Omega_0}$ designates the Laplace-Beltrami operator, which is defined on the surface element $\Omega_0$ through the surface gradient $\nabla_{\Omega_0}$ tangential to the surface. Formally, let ${\bf{n}}_{\Omega_0}$ denote the unit normal to the surface $\Omega_0$ then we define the surface gradient of $\psi$ as $\nabla_{\Omega_0} \psi:=\nabla \psi - (\nabla\psi \cdot {\bf{n}}_{\Omega_0}) {\bf{n}}_{\Omega_0}$ and then the Laplace-Beltrami operator is defined as $\Delta_{\Omega_0} \psi = \nabla_{\Omega_0}\cdot \nabla_{\Omega_0} \psi$.

Note, that in fact, the condition of the fully saturated surface layer is not essential in calculation of the flows over one particle element of the porous media. It is sufficient to presume that the variation of the capillary pressure on the length scale of the particle $\delta\psi$ is negligible, that is $\delta\psi \ll \gamma/\delta_R$. This is usually the case in slow creeping flow conditions in porous media, and in fact, it is a criterion for the use of macroscopic approximation to such flows~\cite{Bear-Book}. In the case when the surface layer is not fully saturated, parameter $\delta_R$ should be interpreted as the effective thickness of the layer filled by the liquid.

At the same time, liquid pressure variation in the bridges is negligible in slow creeping flows in comparison to that in $\Omega_0$. So that, one can assume that
\begin{equation}
\label{Laplace-Beltrami-BCS}
\left. \psi\right|_{\partial \Omega_1}=\psi_1=const, \quad \left. \psi\right|_{\partial \Omega_2}=\psi_2=const,
\end{equation}
which are the boundary conditions to the Laplace-Beltrami Dirichlet boundary value problem. The Dirichlet boundary value problem (\ref{Laplace-Beltrami})-(\ref{Laplace-Beltrami-BCS}) has a unique solution, which, if it is found, allows to calculate the total flux through the particle element
$$
Q_T=\delta_R\frac{\kappa_m}{\mu}\int_{\partial \Omega_1} \frac{\partial \psi}{\partial n}\, dl =  -\delta_R\frac{\kappa_m}{\mu} \int_{\partial \Omega_2} \frac{\partial \psi}{\partial n} \, dl,
$$ 
where $\bf n$ is the normal vector to the domain boundaries $\partial \Omega_{1,2}$ on the surface, $\delta_R$ is the average amplitude of the surface roughness, that is the width of the surface layer conducting the liquid flux and the line integral is taken along a closed curve in $\Omega_0$, for example the boundary $\partial \Omega_1$. 

If the total flux $Q_T$ is determined, one can define the global permeability coefficient of a single particle $K_1$. This can be done, if we assume that the particle has a characteristic size $D$ and so that it can be enclosed in a volume element $V=D^3$ with the characteristic side surface area $D^2$. Then, the effective flux density $Q$ can be represented in terms of $K_1$ (and the total flux $Q_T$)
$$
Q=\frac{Q_T}{D^2}=-\frac{K_1}{\mu}\frac{\psi_2-\psi_1}{D},
$$
if the flow is driven by the constant pressure difference $\psi_2-\psi_1$ applied to the sides of the volume element.

\subsection{Surface permeability of a sphere in the case of azimuthally symmetric domain boundaries.}

Consider now a spherical particle in an azimuthally symmetric case, when the domain boundaries $\partial \Omega_1$ and $\partial \Omega_2$ are oriented at the reflex angle $\alpha=\pi$ and have a circular shape. We use a spherical coordinate system with its origin at the particle centre and the polar angle $\theta$ counted from the axis of symmetry passing through the centre of the circular contour $\partial \Omega_1$. In this case, the Dirichlet boundary value problem (\ref{Laplace-Beltrami})-(\ref{Laplace-Beltrami-BCS}) admits an analytical solution, so that particle permeability can be determined explicitly. Indeed,
problem (\ref{Laplace-Beltrami})-(\ref{Laplace-Beltrami-BCS}), if we assume that the liquid pressure $\psi$ is a function of $\theta$ only and independent of the azimuthal angle, is equivalent to
\begin{equation}
\label{Laplace-Beltrami-symmetric}
\frac{1}{\sin\theta}\frac{\partial }{\partial \theta}\left( \sin\theta \frac{\partial \psi}{\partial \theta}\right)=0, \quad \theta_0 < \theta < \pi-\theta_1,
\end{equation}
with the boundary conditions
\begin{equation}
\label{BCLB-symmetric}
\left. \psi\right|_{\theta=\theta_0}=\psi_1, \quad \left. \psi\right|_{\theta=\pi-\theta_1}=\psi_2.
\end{equation}

The analytic solution to problem (\ref{Laplace-Beltrami-symmetric})-(\ref{BCLB-symmetric}) after applying the boundary conditions can be represented in the following form 
\begin{equation}
\label{Laplace-Beltrami-Analytic}
\psi=\Psi_0 (\psi_2-\psi_1) \ln \left\{\frac{\sin\theta}{\sin\theta_0} \frac{1+\cos\theta_0}{1+\cos\theta}  \right\} +\psi_1,
\end{equation}
where
$$
\Psi_0=\frac{1}{\ln \left\{\frac{\sin\theta_1}{\sin\theta_0} \frac{1+\cos\theta_0}{1-\cos\theta_1}  \right\}}.
$$

One can now calculate the total flux 
$$
Q_T=-\frac{K_1}{\mu} D (\psi_2-\psi_1)=-2\pi \sin\theta_0 \delta_R \frac{k_m}{\mu}\left. \frac{\partial \psi}{\partial \theta}\right|_{\theta=\theta_0}
$$
$$
=-(\psi_2-\psi_1) 2\pi \delta_R \Psi_0 \frac{k_m}{\mu}. 
$$
So that, taking $D=2R$, 
\begin{equation}
\label{PSphere}
K_1=\pi \Psi_0 \frac{\delta_R}{R} k_m.
\end{equation}

One can see that, if we take $\theta_1=\theta_0$, the permeability coefficient $K_1$ is divergent at $\theta_0=\pi/2$, as is expected, when the two contours move closer to each other and, at the same time, their radius $R\sin\theta_0$  increases, that is  
$$
K_1\approx \frac{\delta_R}{2R} \frac{\pi k_m}{(\frac{\pi}{2}-\theta_0)}  \text{ as } \theta_0 \to \frac{\pi}{2}.
$$
In the opposite limit, at $\theta_0=0$, when the two contours move further away from each other and their radius decreases, the permeability coefficient tends to zero, that is
$$
K_1\approx \frac{\delta_R}{2R} \frac{\pi k_m}{|\ln\theta_0 |}  \text{ as } \theta_0 \to 0.
$$
Parametrically, the coefficient of permeability (\ref{PSphere}) is inversely proportional to the particle radius $R$, so that larger particles create stronger resistance to the flow. Noticeably, the coefficient demonstrates strong dependence on the surface layer thickness $\delta_R$, that is $K_1\propto \delta_R^3$ since it is anticipated that $k_m\propto \delta_R^2$, so that evaluation of this parameter in applications is crucial for the accurate estimates of the liquid dispersion rates. 

How does the result affect the super-fast diffusion model (\ref{Superfast-1}), and basically how can it be incorporated into the main diffusion equation? If we approximate the permeability coefficient $K$ by $K_1$ obtained in the azimuthally symmetric case at $\theta_1=\theta_0$, and, using an approximate relationship between the radius of curvature $R\sin\theta_0$ of the boundary contour $\partial \Omega_1$ and the pendular ring volume~\cite{Herminghaus-2005}, one can show
$$
\sin^2\theta_0 \approx \sqrt{s-s_0} 
$$
and at $\theta_0\ll 1$ or $(s-s_0)\ll 1$
\begin{equation}
\label{GPSphere}
K\approx 2 \frac{\delta_R}{R} \frac{\pi k_m}{|\ln(s-s_0)|}.
\end{equation}
As one can see from (\ref{GPSphere}), the distinctive particle shape results in logarithmic correction to the main non-linear superfast-diffusion coefficient $D(s)=\frac{D_0}{(s-s_0)^{3/2}}$, such that 
$$D(s)\propto \frac{1}{|\ln(s-s_0)|(s-s_0)^{3/2}}.$$ Apparently, the correction will mitigate to some extent the divergent nature of the dispersion at the very small saturation levels $s\approx s_0$, smoothing out the characteristic dispersion curves. 

\begin{figure}[ht!]
\begin{center}
\includegraphics[trim=-1.5cm -3.cm 0cm 0.5cm,width=\columnwidth]{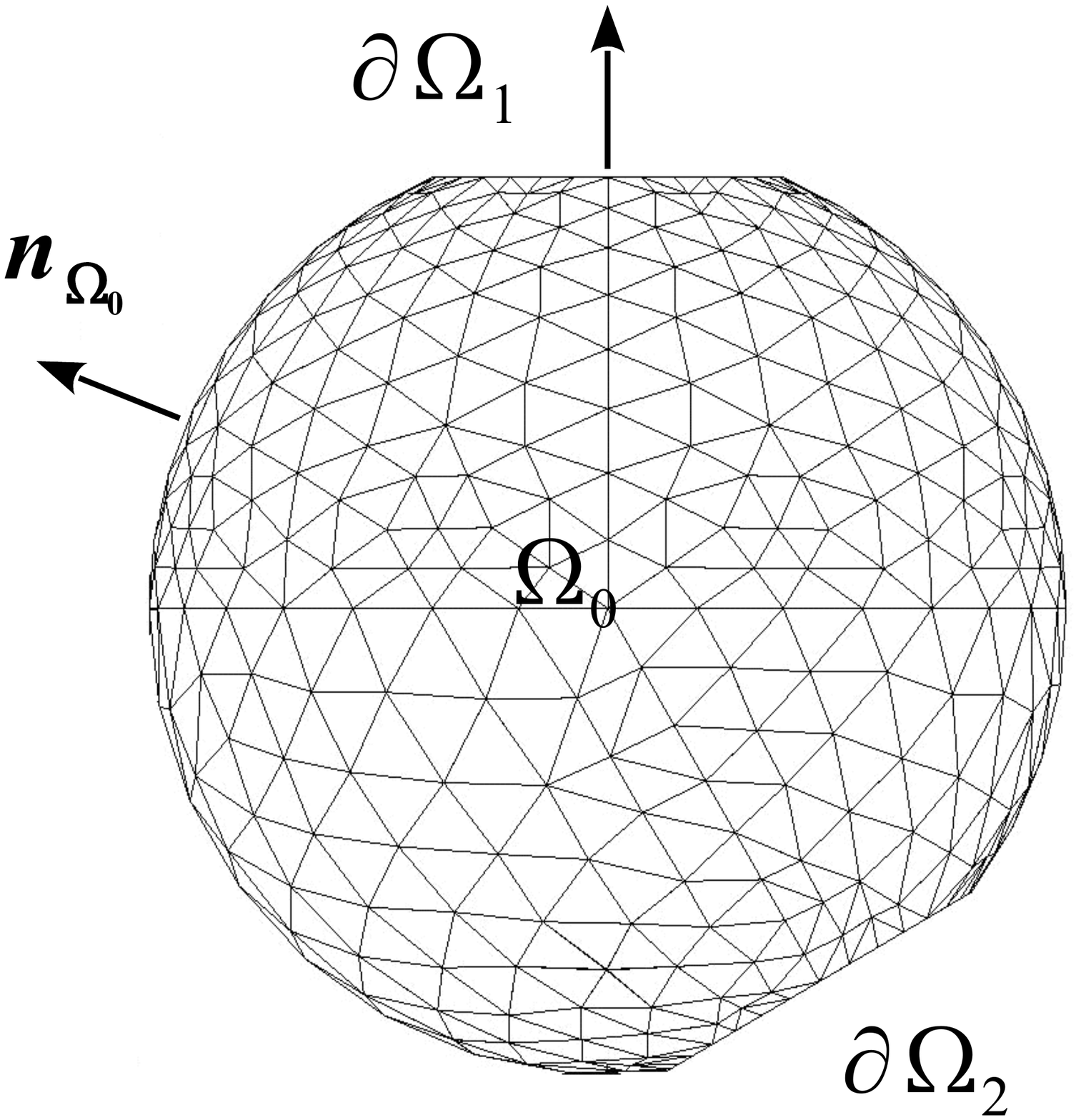}
\end{center}
\caption{Illustration of the triangular tessellation of the truncated spherical surface domain $\Omega_0$ with a normal vector ${\bf n}_{\Omega_0}$ at $\alpha=5\pi/6$ and $\theta_0=\theta_1=\pi/8$.} 
\label{Fig3}
\end{figure}

\begin{figure}[ht!]
\begin{center}
\includegraphics[trim=-1.5cm 1.cm 1cm -0.5cm,width=\columnwidth]{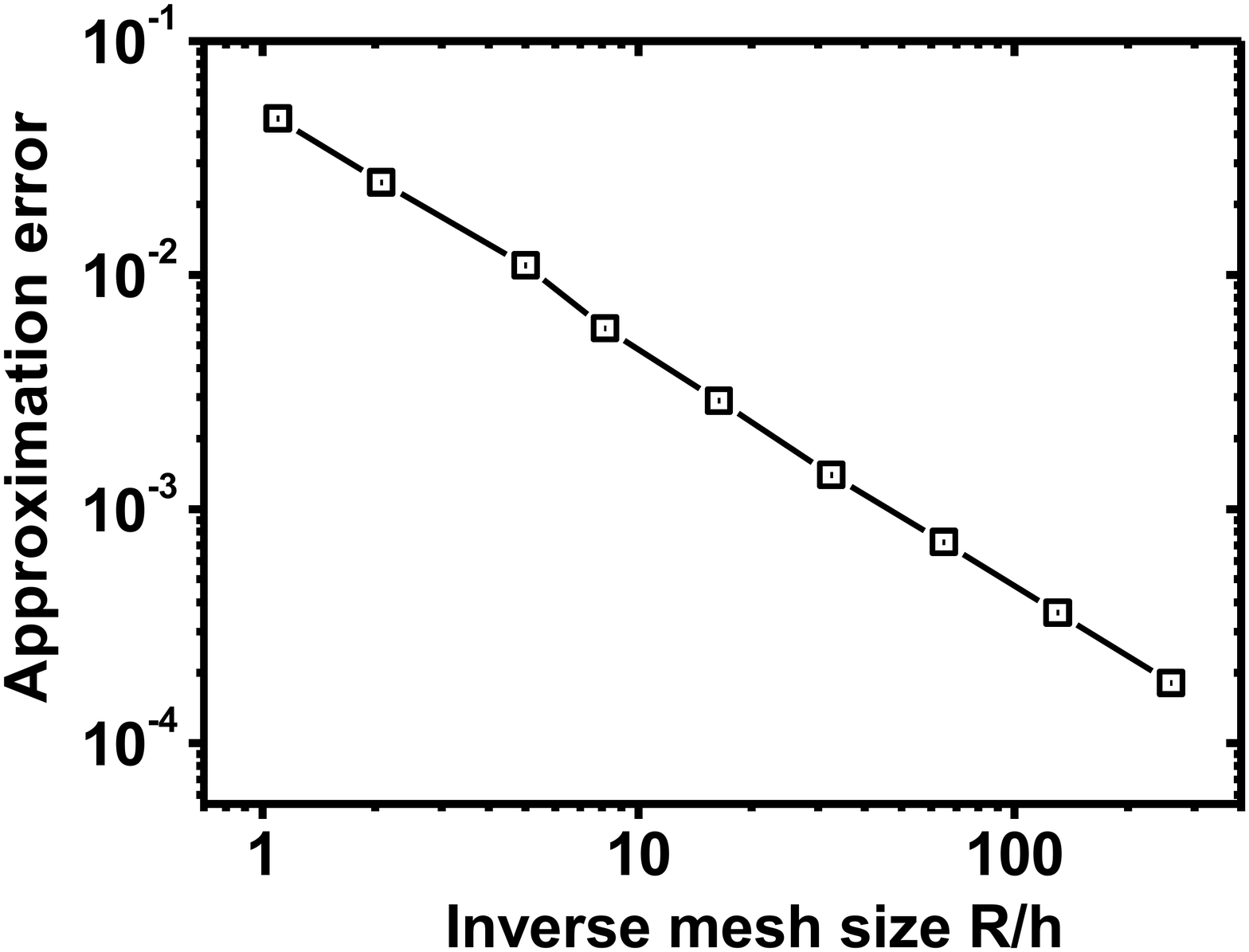}
\end{center}
\caption{Verification of the numerical scheme in the azimuthally symmetric scenario. We plot the inverse mesh size against the  error measured in the energy norm \cite{Dziuk1988}. We observe the rate of convergence proven in \cite{Dziuk1988} verifying that, asymptotically, the numerical approximation converges to the exact solution.} 
\label{Fig-Error}
\end{figure}

Before we proceed to a general case, this would be instructive to consider, in qualitative terms, how specific is the permeability of spherical particles. We now compare coefficient of permeability (\ref{PSphere}) with the permeability of a cylinder of radius $R \sin\theta_0$ and length $2R$ with the same surface layer of thickness $\delta_R$. Such an element was often used in simple estimations of permeability in porous media~\cite{Koorevaar-1983}. It is not difficult to calculate the total flux through this element when there is a constant pressure difference $(\psi_2-\psi_1)$ applied to its ends
$$
Q_T=-(\psi_2-\psi_1) \frac{k_m}{\mu} \pi  \sin\theta_0 \delta_R =-2R \frac{K_c}{\mu} (\psi_2-\psi_1),
$$
so that
$$
K_c=\pi k_m \sin\theta_0 \frac{\delta_R}{2R}\propto (s-s_0)^{1/4},
$$
where $K_c$ is the effective permeability of the cylindrical element.

One can observe, that in contrast to the case of spherical elements, the cylindrical approximation provides completely different correction to the non-linear coefficient of diffusion, if we presume similar scaling $\sin^2\theta_0 \approx \sqrt{s-s_0}$. Consider now a general case.

\subsection{Surface permeability of a sphere in the case of arbitrary oriented boundaries.} 
In the arbitrary case, when $\alpha\neq \pi$, the Dirichlet boundary value problem (\ref{Laplace-Beltrami})-(\ref{Laplace-Beltrami-BCS}) does not possess known explicit solutions, so  we make use of a classical surface finite element technique introduced in \cite{Dziuk1988}. See also \cite{Dziuk2013} for an in depth review of state of the art innovations and uses pertaining to this class of method. Using this method we are able to numerically investigate the total flux and hence the permeability of the particle. 

\begin{figure}[ht!]
\begin{center}
\includegraphics[trim=-1.5cm 1.cm 1cm -0.5cm,width=\columnwidth]{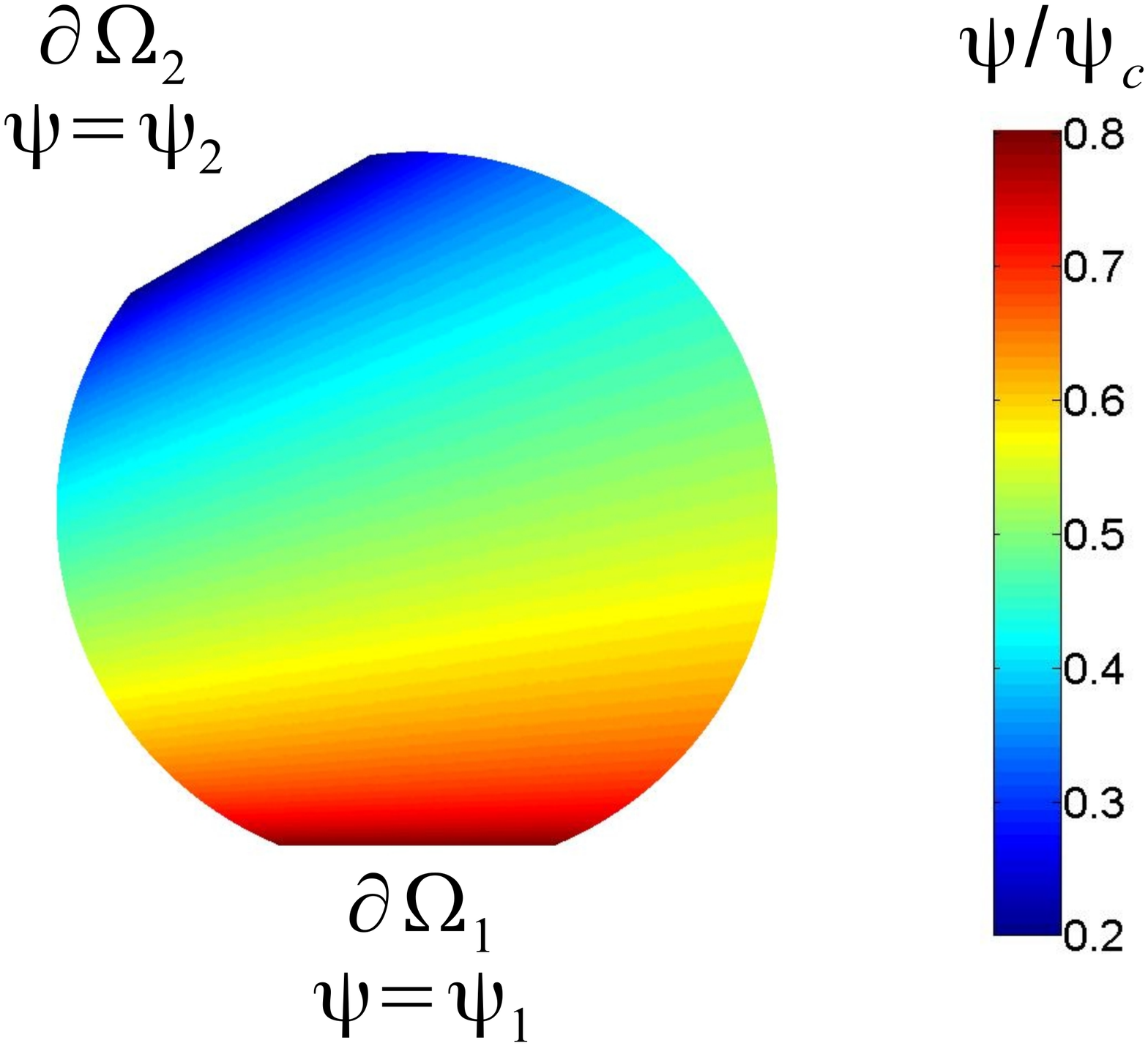}
\end{center}
\caption{Distribution of non-dimensional pressure $\psi/\psi_c$ ($\psi_c=2\gamma/R$) on a truncated unit sphere with identical circular boundary contours at $\psi_1/\psi_c=0.8$, $\psi_2/\psi_c=0.2$, $\theta_1=\theta_0=\pi/8$ and $\alpha=5\pi/6$. The colour bar indicates the value of non-dimensional pressure $\psi/\psi_c$.} 
\label{Fig4}
\end{figure}

We begin by approximating the truncated surface element with a piecewise linear approximation through triangular elements, see Fig. \ref{Fig3} for an example. In this setting, we are approximating the geometry with a polygon. This inherently introduces an error through the approximation of the geometry. It is, however, well understood appearing as a 'variational crime' \cite{Dziuk2013}. We then discretise the Laplace-Beltrami operator over the polygon using piecewise linear finite elements. To test our numerical model we examine the azimuthally symmetric case, where the exact solution is known and given in (\ref{Laplace-Beltrami-Analytic}). We then check convergence of the finite element approximation to (\ref{Laplace-Beltrami-Analytic}). The results are shown in Fig. \ref{Fig-Error}.

We make use of the numerical model generated to examine the dependency of the total flux, and hence the permeability of the truncated spherical element as a function of the tilt angle $\alpha$, that is the position of the boundaries on the sphere at fixed values of the capillary pressure $\psi_1$ and $\psi_2$. As in the azimuthally symmetric case, without much loss of generality, we consider circular boundaries. The size of the boundary contour, that is its radius $R\sin\theta_0$ (or $R\sin\theta_1$), will be characterized by the polar angle $\theta_0$ (or $\theta_1$) counted from the axis of symmetry of each contour and the particle radius $R$.

\subsection{Results of numerical analysis and discussion}
\begin{figure}[ht!]
\begin{center}
\includegraphics[trim=-1.5cm 1.cm 1cm -0.5cm,width=\columnwidth]{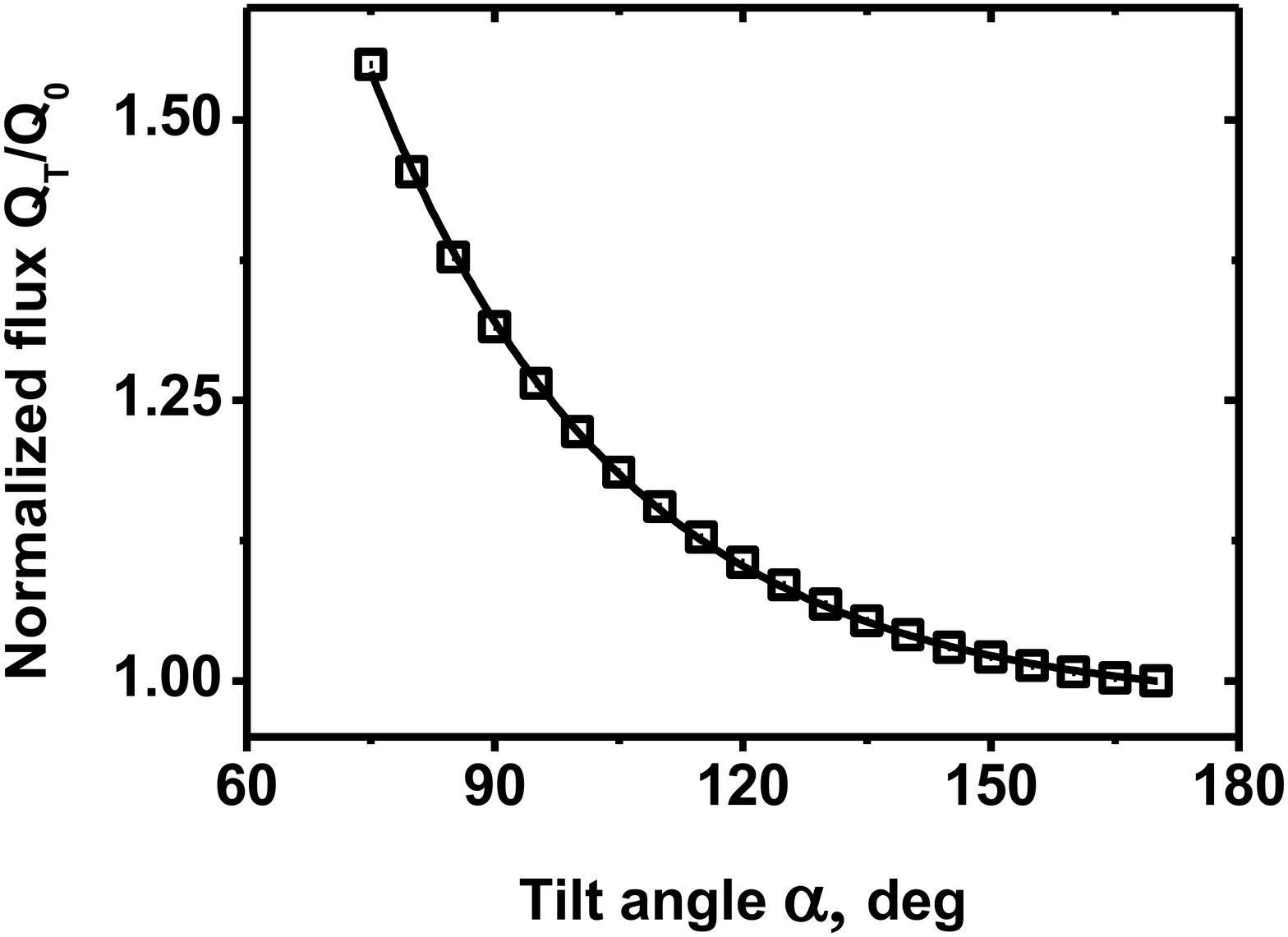}
\end{center}
\caption{Non-dimensional total flux $Q_T/Q_0$ as a function of the tilt angle $\alpha$ at $\theta_0=\theta_1=\pi/8$ and fixed values of the capillary pressure at the boundaries $\psi_1=0.8$ and $\psi_2=0.2$. Here $Q_0$ is the total flux value at $\alpha=\pi$. The numerical result obtained at high resolution (maximum mesh size $h/R\approx 0.003$) is shown by symbols and the solid line is the best fit to the data $Q_T/Q_0=B_0+B_1\exp(-(\alpha-\alpha_0)/\Delta_{\alpha})$ at $B_0=0.98$, $B_1=1.5$, $\alpha_0=46^{\circ}$ and $\Delta_{\alpha}=30^{\circ}$. The approximation error is about the symbol size.} 
\label{Fig5}
\end{figure}

The distribution of pressure on the spherical surface is illustrated in Fig. \ref{Fig4}, while the typical total flux dependence on the tilt angle $\alpha$ is presented in Fig. \ref{Fig5} at $\theta_0=\theta_1$ and at fixed values of $\psi_1$ and $\psi_2$. The distribution of pressure demonstrates relatively smooth variations in the range bounded by the prescribed boundary values, such that, as is expected in a diffusion problem, $\psi_2 \le \psi \le \psi_1$. The value of the total liquid flux $Q_T$ through the spherical element decreases when the tilt angle increases and the boundary contours move further away from each other. At the same time, one readily observes, Fig. \ref{Fig5}, that at relatively large tilt angles, close to the reflex angle in the azimuthal symmetrical case, the total flux value and hence permeability of the surface elements, is close to that predicted on the basis of the azimuthally symmetric solution (\ref{PSphere}). This implies that the analytical result (\ref{PSphere}) and (\ref{GPSphere}) can be used in practical applications to obtain first order corrections to the effective non-linear coefficient of dispersion in the super-fast diffusion model. One may notice that even at small tilt angles, when the two boundaries are located close to each other, one can still approximate coefficient of permeability with the accuracy of $50\,\%$. We have verified numerically that in the general case the permeability coefficient of the particles demonstrates the same trends with variations of parameters $\theta_0$ and $\theta_1$ as in the azimuthally symmetric case.

\begin{figure}[ht!]
\begin{center}
\includegraphics[trim=-1.5cm 1.cm 1cm -0.5cm,width=\columnwidth]{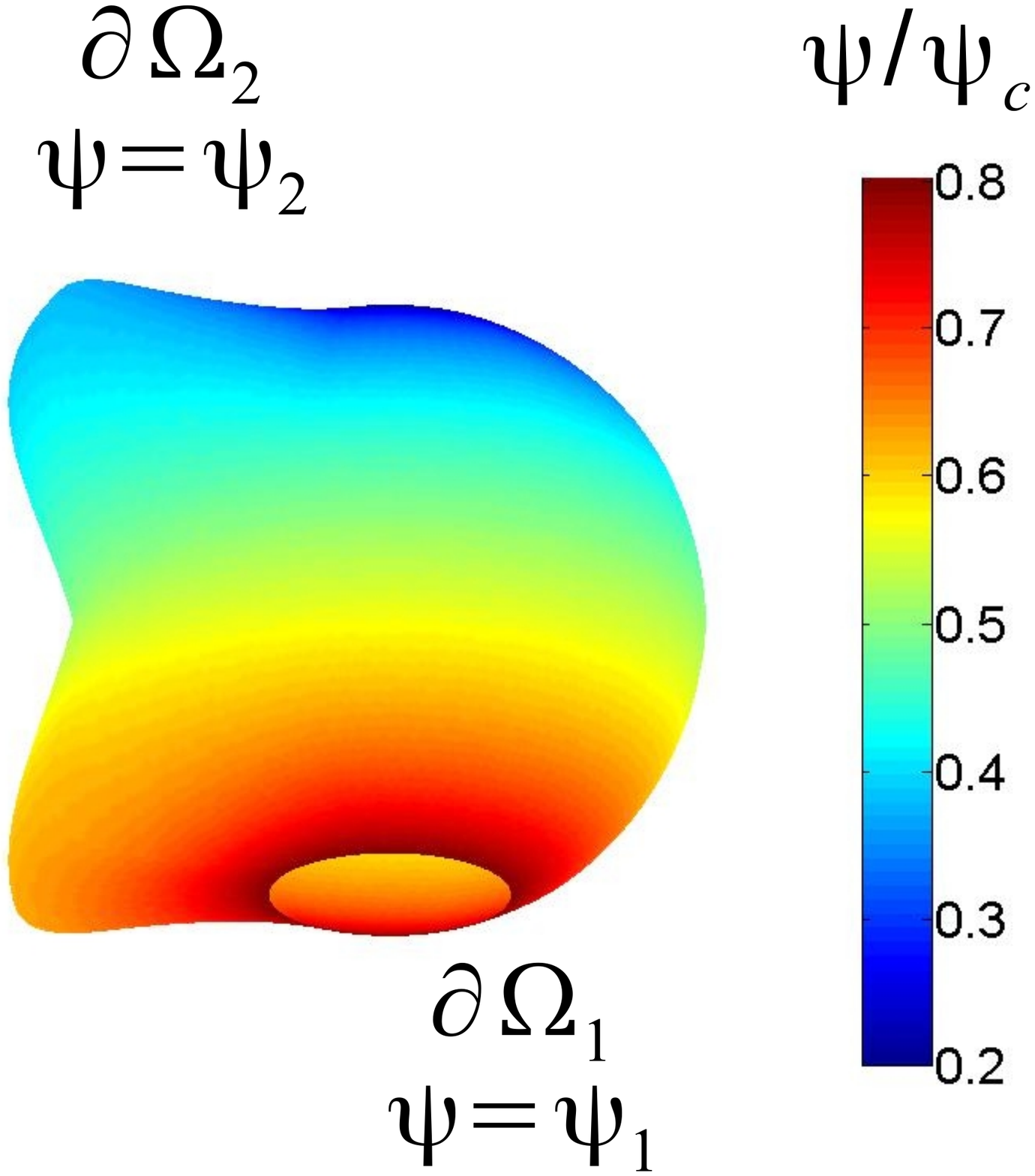}
\end{center}
\caption{Particle shape and distribution of non-dimensional pressure $\psi/\psi_c$ ($\psi_c=2\gamma/R$) on an arbitrary truncated surface with identical circular boundary contours at $\psi_1/\psi_c=0.8$, $\psi_2/\psi_c=0.2$, $\theta_1=\theta_0=\pi/8$ and $\alpha=\pi$. The colour bar indicates the value of non-dimensional pressure $\psi/\psi_c$. Non-dimensional total flux $Q_T/Q_0\approx 0.86$, where $Q_0$ is the total flux value at $\alpha=\pi$ through the original truncated sphere used to generate the arbitrary surface shape.} 
\label{Fig6}
\end{figure}

\begin{figure}[ht!]
\begin{center}
\includegraphics[trim=-1.5cm 1.cm 1cm -0.5cm,width=\columnwidth]{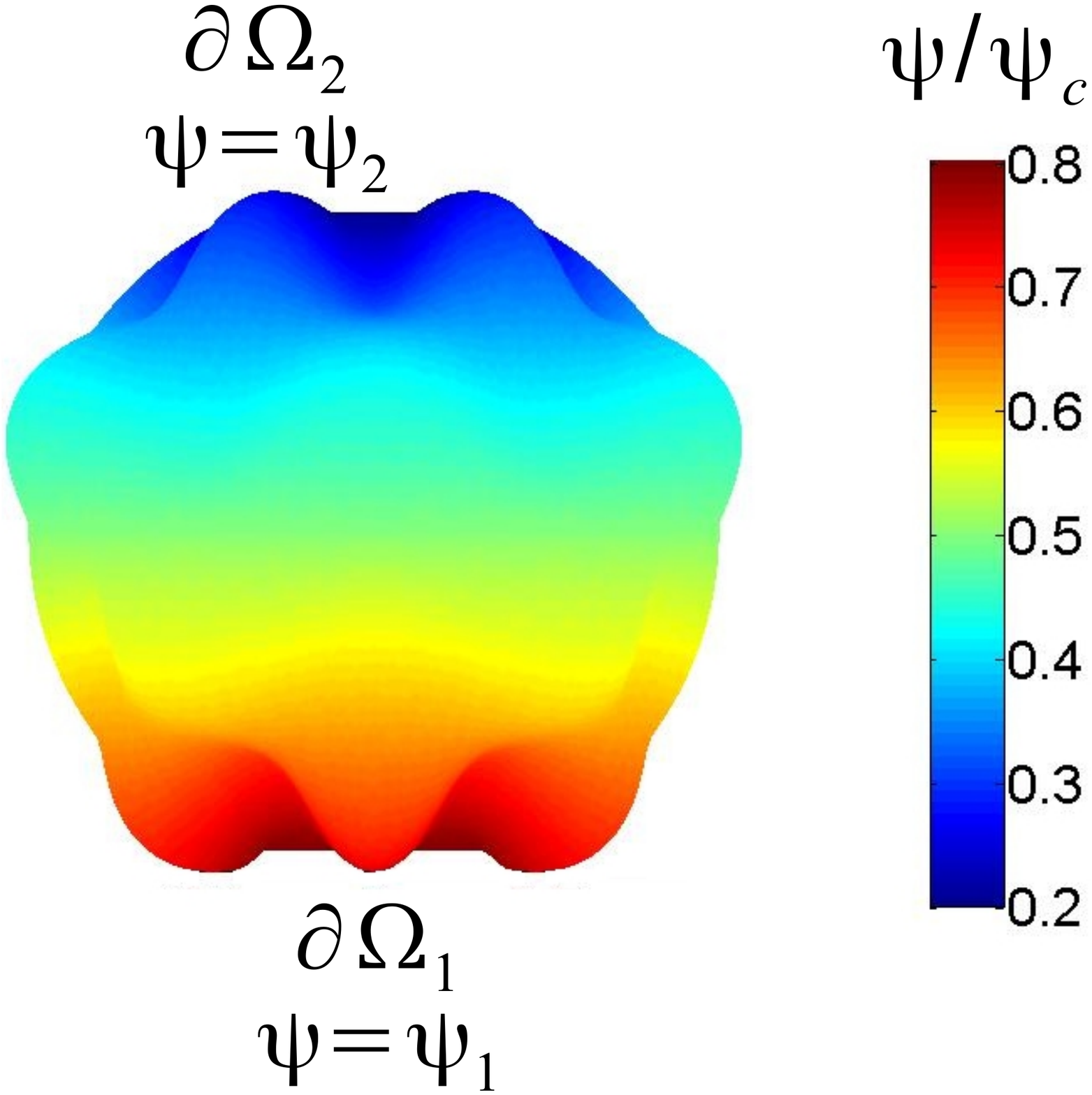}
\end{center}
\caption{Particle shape and distribution of non-dimensional pressure $\psi/\psi_c$ ($\psi_c=2\gamma/R$) on the truncated surface $r(\theta,\phi)=R(1+A_s\cos m\theta\cos n\phi)$ at $m=n=5$ and $A_s=0.15$ with identical circular boundary contours at $\psi_1/\psi_c=0.8$, $\psi_2/\psi_c=0.2$, $\theta_1=\theta_0=\pi/8$ and $\alpha=\pi$. The colour bar indicates the value of non-dimensional pressure $\psi/\psi_c$. Non-dimensional total flux $Q_T/Q_0\approx 0.95$, where $Q_0$ is the total flux value at $\alpha=\pi$ through the original truncated sphere used to generate the perturbed surface shape.} 
\label{Fig7}
\end{figure}

\begin{figure}[ht!]
\begin{center}
\includegraphics[trim=-1.5cm 1.cm 1cm -0.5cm,width=\columnwidth]{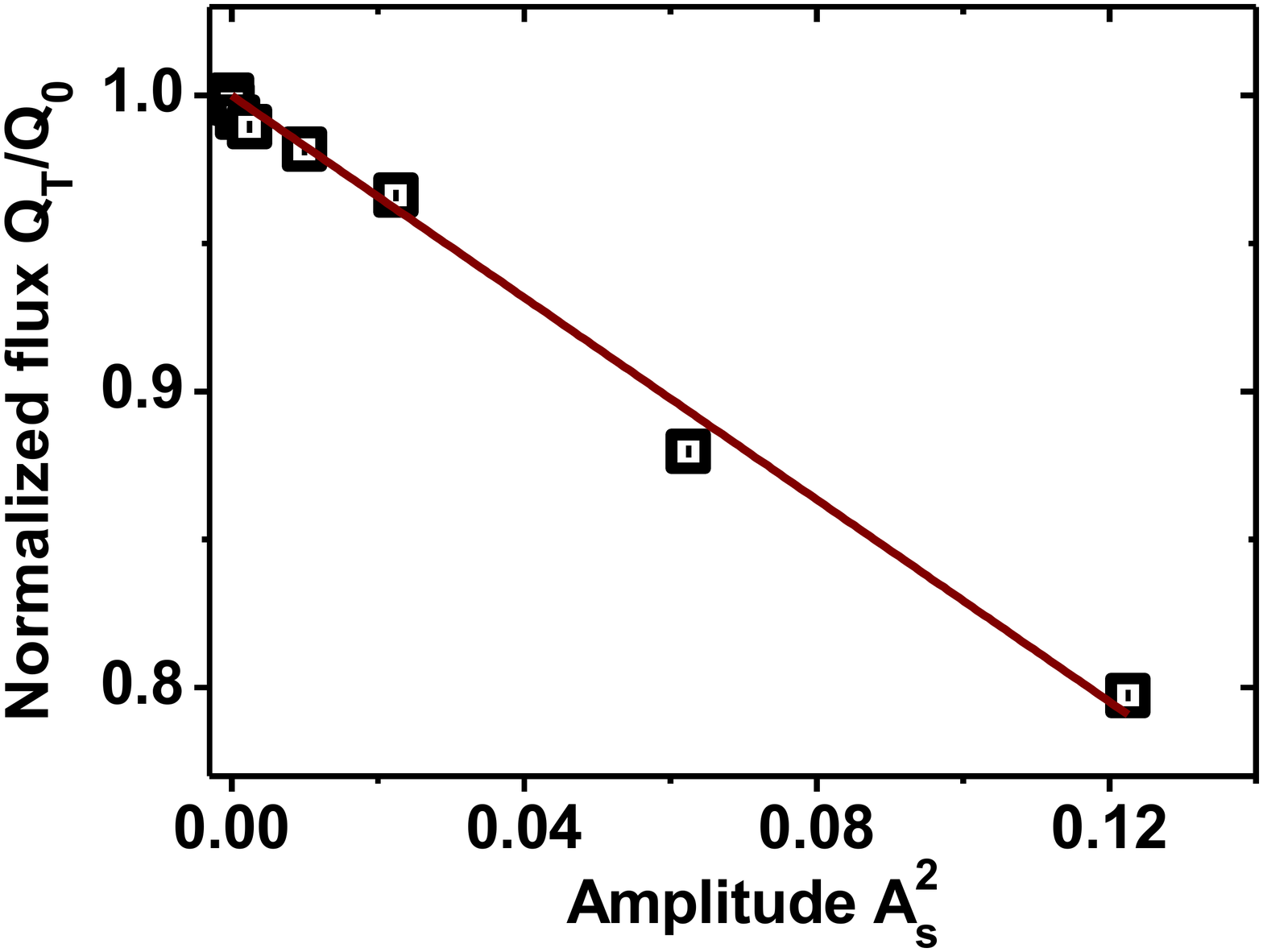}
\end{center}
\caption{Non-dimensional total flux $Q_T/Q_0$ as a function of the shape perturbation amplitude $A_s^2$ at fixed values of $\psi_1/\psi_c=0.8$ and $\psi_2/\psi_c=0.2$, $\alpha=\pi$, $m=n=5$ and $\theta_0=\theta_1=\pi/8$. Here $Q_0$ is the total flux value through the unperturbed spherical element in similar conditions. The numerical results obtained at medium resolution (maximum mesh size $h/R\approx 0.03$) are shown by symbols and the solid line is the best fit $Q_T/Q_0=1-C_s A_s^2$ at $C_s=1.7$. The approximation error is about the symbol size.} 
\label{Fig8}
\end{figure}

\subsection{Arbitrary particle shapes}
Even low dispersed sand samples consist of grain particles, which are only approximately spherical~\cite{Alshibli2004}. Therefore, we consider arbitrary surface elements obtained by perturbations of a sphere preserving surface smoothness. Based on our methodology, we examine numerical solutions to the Laplace-Beltrami Dirichlet boundary value problem (\ref{Laplace-Beltrami}) set on such perturbed particle surfaces to calculate the total volumetric flux, which is the measure of the surface permeability. To separate the effects of the particle shape from the effects of the boundary shape on the particle surface permeability and for the sake of comparison with the permeability of spherical particles, we consider circular boundary contours oriented to each other as in the azimuthally symmetric case, Fig. \ref{Fig6}. The size of the boundary contour, that is its radius $R\sin\theta_0$ (or $R\sin\theta_1$), will be characterized by the polar angle $\theta_0$ (or $\theta_1$) counted from the axis of symmetry of each contour and the radius of the sphere used to obtain the perturbed surface element $R$. The first particle shape, we have examined, is shown in Fig. \ref{Fig6} with the distribution of the liquid pressure indicated by the colour map. For the sake of comparison, we have chosen the same boundary conditions as in the case of spherical shapes, that is $\psi_1/\psi_c=0.8$ and $\psi_2/\psi_c=0.2$, with the same contour sizes, that is $\theta_1=\theta_0=\pi/8$ oriented at $\alpha=\pi$.  As is expected, the total volumetric flux, in this case $Q_T$, is reduced in comparison with that, $Q_0$, through the spherical particle shape $Q_T/Q_0\approx 0.86$, since some pathways connecting two boundary contours became much longer, as one can see from Fig. \ref{Fig6}. Despite, at first glance, strong variations of the original spherical shape, the observed effect is not dramatic and is on the scale of the change of the surface area demonstrating that the spherical shape provides a good approximation in general to obtain estimates of the surface permeability. Indeed, the total increase of the surface area due to the perturbation was $S_a/S_0\approx 1.2$, where $S_0=4\pi R^2\cos\theta_0$ is the surface area of the original truncated spherical particle, so that the characteristic size of the particle calculated via $R_a=R \sqrt{S_a/S_0}\approx 1.1\, R$. We note though that the actual parameter defining the particle permeability is expected to be an effective length of the pathways connecting the boundary contours.        

In general, effective pathway length scale is not so easy to estimate, therefore, to understand the role of this effective parameter, consider now specific systematic changes of the original spherical shape of radius $R$ via the transformation of the form 
\begin{equation} r(\theta,\phi)=R\, (1+A_s\cos m\theta\, \cos n\phi)
\label{trans}
\end{equation} 
where $\theta$ and $\phi$ are the polar and azimuthal angles of the spherical coordinate system.

The obtained surface profile is demonstrated in Fig. \ref{Fig7} at $A_s=0.15$ and $m=n=5$.  As in the previous case, the boundary contours are circular, identical ($\theta_1=\theta_0$) and are not perturbed. The smoothness of the perturbed surface shape was achieved via a spline approximation at the boundary contours during the mesh generation and further refinement of the mesh. In this procedure, a smooth surface profile is created with two small boundary regions, which are not exactly described by the transformation (\ref{trans}). In what follows, we fix parameters of the perturbation transformation $m=n=5$ and consider only variations of the amplitude $A_s$. Variation of the total flux $Q_T$ through such elements with the amplitude of the perturbation $A_s$ is shown in Fig. \ref{Fig8}.  

The characteristic arc length $L_p$ of the perturbed shape can be estimated by means of 
$$\frac{L_p}{(\pi-2\theta_0) R}\approx 1+\frac{A_s^2 m^2}{8} $$ 
at $A_s^2 m^2/8\ll 1$ and $\theta_0=\theta_1$. The estimate follows from the definition of $L_p$ along the meridian line ($\phi=const$) taking into account that $A_s^2 m^2/8\ll 1$ and applying averaging in the azimuthal direction, that is over $\phi$,
$$
L_p=\int_{\theta_0}^{\pi-\theta_0} \sqrt{1+\frac{1}{R^2}\left(\frac{\partial r}{\partial \theta} \right)^2}\, R\, d\theta = 
$$
$$
\int_{\theta_0}^{\pi-\theta_0} \sqrt{1+A_s^2 m^2 \sin^2 m\theta\,\cos^2 n\phi}\, R\, d\theta = 
$$
$$
\int_{\theta_0}^{\pi-\theta_0} \sqrt{1+\frac{A_s^2 m^2}{4} (1-\cos 2m\theta)(1+\cos 2 n\phi) }\, R\, d\theta\approx 
$$ 
$$
\int_{\theta_0}^{\pi-\theta_0} \left\{ 1+\frac{A_s^2 m^2}{8} (1-\cos 2m\theta)(1+\cos 2 n\phi) \right\}\, R\, d\theta.
$$
That is after averaging over the azimuthal angle and neglecting contribution of the term of the order of $\sin\theta_0 /2m\ll 1$ 
$$
L_p \approx \int_{\theta_0}^{\pi-\theta_0} \left\{ 1+\frac{A_s^2 m^2}{8} (1-\cos 2m\theta)  \right\}\, R\, d\theta 
$$
and
$$
L_p \approx  (\pi - 2\theta_0)\, R\, \left\{ 1+\frac{A_s^2 m^2}{8}\right\}.
$$

Since the total volumetric flux is expected to be proportional to the pressure gradient, one can anticipate that its dependence on the effective arc length would follow $Q_T/Q_0\approx \frac{(\pi-2\theta_0) R}{L_p} \approx 1 - \frac{A_s^2 m^2}{8}$. As one can see, Fig. \ref{Fig8}, the numerically calculated total flux dependence does follow the trend suggested by scaling of the arc length $L_p$, the match though is not perfect. We found from the best fit $Q_T/Q_0=1-C_s A_s^2$, Fig. \ref{Fig8}, that $C_s=1.7$, while the value $C_s\approx 3.1$ would be expected. This implies that the surface diffusion over uneven landscapes is a slightly more complex phenomenon than that one would expect from the simple scaling suggested by the effective pathways length. We note, in that respect, that the methodology and the numerical treatment of the Laplace-Beltrami problem developed are particularly indispensable, where there is no simple way of estimating the effective parameter $L_p$, for example over strongly heterogeneous surface profiles with large areas inaccessible to the liquid flow.

\section{Conclusions}
We have demonstrated how the permeability coefficient of constituent elements of a porous matrix can be estimated on the basis of a solution to the Laplace-Beltrami problem using, as an example, truncated spherical particles with arbitrary oriented boundaries and perturbed spherical shapes. In the azimuthally symmetric case, we obtained an observable analytical solution, which has been incorporated into the macroscopic super-fast dispersion model to calculate a correction to the effective non-linear coefficient of diffusion. We have shown, that in the case of arbitrary oriented boundaries and perturbed spherical shapes, the analytical solutions provide a reasonable approximation in the general case. The analytical, (\ref{PSphere}) and (\ref{GPSphere}), and numerical solutions are the main results of our paper. The methodology developed in our study can be used in practical applications involving more sophisticated shapes of constituent elements and their compositions. This will be the subject of future studies.

\bigskip\bigskip
PS was supported through the Royal Thai Government scholarship. TP was partially supported through the EPSRC grant EP/P000835/1.

\end{document}